\documentclass[preprint,review,12pt]{iopart}

\usepackage{graphicx}
 \usepackage[dvips]{color}

\begin{document}
 
\title{Iron oxide doped boron nitride nanotubes: structural and magnetic properties} 

\author{ Ronaldo J. C. Batista, Alan B. de Oliveira, and Daniel L. Rocco }
\address{Departamento de F\'{\i}sica, 
Universidade Federal de Ouro Preto, Ouro Preto, MG, Brazil, 35400-000.}

\ead{batista.rjc@iceb.ufop.br}

\begin{abstract}

    A first-principles formalism is employed to investigate the interaction of iron oxide (FeO) with a boron nitride (BN) nanotube.   
The stable structure of the FeO-nanotube % complexe 
has Fe atoms binding  N atoms, with bond length of roughly $\sim$2.1 \AA, and binding between O and B atoms, 
with bond length of 1.55 \AA. In case of small FeO concentrations, the total magnetic moment is (4$\mu_{Bohr}$ ) 
times the number of Fe atoms in the unit cell and it is energetically favorable to FeO units to aggregate rather than 
randomly bind to the tube. As a larger FeO concentration case, we study a BN nanotube fully covered by a single layer of FeO. 
We found that such a structure has square FeO lattice with Fe-O bond length of 2.11 \AA, similar to that of FeO bulk, 
and total magnetic moment of 3.94$\mu_{Bohr}$ per Fe atom. Consistently with experimental results, the FeO covered 
nanotube is a semi-half-metal which can become a half-metal if a small change in the Fermi level is induced. 
Such a structure may be important in the spintronics context.

\end{abstract}

\maketitle

\section{Introduction}

In the last decades near one dimensional systems, such as nanowires \cite{nanowire:06}, 
nanotubes \cite{nanotubefeo:04} and organic molecules \cite{boca07},
has attracted attention of the scientific community. Great part of 
such an interest is due their potential for applications in micro- and nano-technological devices. 
In particular, ferromagnetic nanotubes may be of scientific and industrial interest because they
potentially serve as tunable fluidic channels for tiny magnetic particles,
data storage devices in nanocircuits, and scanning tips for magnetic force
microscope \cite{nanotubefeo:04,s03}. Moreover, it has been recently shown that ferromagnetic iron
oxide nanotubes can be used as building blocks for spin-based devices \cite{xyhz05}.

The possibility of delivering information based on the electron spin,
besides on its charge, was a technology breakthrough.
The addition of an extra degree of freedom -- which comes with the electron spin --
to a conventional charge-based device improves the original quantity of information which can be 
manipulated in some orders of magnitude. 
Important applications of spintronics include magnetic field sensors, read heads for hard drives, 
galvanic isolators, and magnetoresistive random access memory to cite a few \cite{wabd01,p98}. 
The controlling of the electron spins is a crucial matter 
for spin-based electronics.  In this sense,  the so-called half-metals, i.e., materials which 
have a metal-like behaviour for one spin channel and are semiconductor for
the other one,  become essentially important for developing spin-based electronics
since they may work as spin filters \cite{xyhz05}.
The first evidence of half-metallic behavior is due to the theoretical work of de Groot \emph{et al.}, 
in which a Mn-based Heusler alloy was studied \cite{gmeb83}.
Recently, Shan and collaborators \cite{sswk09} have observed that the half-metallic Co-based full-Heusler alloy 
Co$_2$FeAl$_{0.5}$Si$_{0.5}$ exhibits the highest effective spin polarization at 300 K and has the weakest temperature dependence 
of spin polarization among all know half-metals. 
This later characteristic become very important since spintronics devices are commonly used at the room temperature. 

Some ferromagnetic oxides also present half-metallic behavior. We can cite,
for instance, the CrO$_2$ \cite{wwmb00} and some members of the mixed valence perovskites 
(manganite), as  the La$_{2/3}$Sr$_{1/3}$MnO$_3$ \cite{lbbj06}. 
In spintronics context, one important ferromagnetic oxide is the Fe$_3$O$_4$ (ferrite)\cite{xyhz05}. 
Such an oxide presents high Curie temperature ($\sim$ 858 K) and half-metallicity, which was predicted by Zhang and Satpathy \cite{magnetite1} and later confirmed by Liao \emph{et al.} \cite{llxz06}. 
Recently, it has been reported that Co ferrite fabricated by thermal oxidation presents spin filter 
efficiencies of 44 \% and 4.3 \% at 10 K and room temperature \cite{apl:2010}, respectively, 
showing that such materials can be, in principle, used in spin filter devices at room temperatures. In addition,
one of the greatest advantages of oxides is their resistance to attack of oxygen atoms present in the atmosphere.

Boron nitride (BN) nanotubes are insulators independently of their chiralities.
Besides, they are more chemically inert than carbon nanotubes with as much as their high thermal stability.
Thus, it is reasonable to consider BN nanotube-based systems as
convenient candidates for application as building blocks in spin-filter devices \cite{yzl04,xyhz05,mylg09}.
In this direction, Xiang and coworkers have studied 
Ni encapsulated BN nanotubes by means of \emph{ab initio} calculations. They 
have found that such a material presents semi-half-metallic behaviour, 
which can become half-metallic after doping electrons more than 1.4 $e$ per unit cell \cite{xyhz05}. 
Another BN nanotube-based half-metal was reported by Min and collaborators \cite{mylg09}, who performed
first-principles calculations on AuV(Cr) quantum wires adsorbed on a (5,5) BN nanotube. 

Hybrid structures composed of BN nanotubes and ferromagnetic oxides may present the previously described advantages of 
both components. Among the ferromagnetic oxides, the iron oxides are particularly interesting since: 
(i) Fe, O, FeO and Fe$_{2}$O$_{3}$ are often present in the synthesis process of boron nitride nanotubes 
\cite{wkyf05,lwky08,mkigy07,kydg04}, moreover, hybrid structures composed of BN doped with Fe or FeO have 
been theoretically predicted \cite{bmc07,wz06} and synthesized \cite{yt01,os01}; 
(ii) iron oxides are important in the spintronics context. It has been shown that Fe$_{2}$O$_{3}$ nanotubes are efficient 
half-metals which can form the basis for spin-filter devices \cite{xyhz05,llxz06}. 

In this work, we employ first-principles calculations to investigate the interaction of iron oxide, FeO, with a (10,0) BN nanotube. 
We found that FeO molecules 
and BN nanotubes may form stable structures. The binding energy of a FeO molecule and the tube 
is 1.7~eV, which is 1.1~eV larger than the binding energy of the tube with an iron atom \cite{wz06}. 
We also found that it is energetically preferable for FeO molecules to aggregate rather than randomly bind to the tube. 
In case of small FeO concentrations, the total magnetic moment is (4$\mu_{Bohr}$ ) times the number of Fe atoms in the unit cell. 
As a larger FeO concentration case, we investigate the BN nanotube fully covered with a single layer of FeO. 
Such an structure presents a total magnetic moment of 3.94$\mu_{Bohr}$ per Fe atom. Consistently with
the experimental results of Liao \emph{et al.} \cite{llxz06}, the FeO covered nanotube is a semi-half-metal which can 
theoretically become half-metal if a small change in the Fermi level is induced ($\sim$0.11 eV). Due to the high value of binding 
energy between FeO molecules and the BN nanotube (1.7~eV), we expect that no special condition and/or artifacts 
(such as flatting the nanotube in order to increase its reactivity \cite{zg09,amc08}) should be required to obtain the FeO 
doped BN nanotubes. Therefore, such a system is an interesting candidate to be experimentally synthesized.

\section{Methodology}

Our  first-principles methodology is based
on the Density Functional Theory (DFT) as implemented
in the SIESTA program \cite{siesta}. We used the Generalized Gradient
Approximation (GGA) as parameterized in the Perdew-Burke-Ernzerhof scheme (PBE) \cite{pbe96} for 
the exchange-correlation functional. The ionic core potentials were represented
by norm-conserving scalar relativistic Troullier-Martins \cite{tm91} pseudopotentials
in Kleinman-Bylander nonlocal form \cite{kb82}. 
The fineness of the real-space grid integration was defined by a minimal energy
cutoff of 120~Ry \cite{ms92}. The geometries were fully optimized until all the 
forces components become smaller than 0.04~eV/\AA. The solutions of the Kohn-Sham equations 
were expanded in a basis set composed of numerical atomic orbitals of finite range \cite{jpsa01}. 
To describe the Fe and O orbitals it was employed two functions per angular momentum plus polarization 
orbitals, the DZP basis set \cite{jpsa01}.
In the case of N and B orbitals it was employed two functions per angular momentum, the DZ basis set.

In order test the reliability of our calculations, we have compared results for the $B_{36}N_{36}$ fullerene, 
the cubic BN, the FeO molecule, and the FeO bulk using the methods described above
with other theoretical and experimental results. 

Regarding the properties of the $B_{36}N_{36}$ fullerene, we have found
the mean distance of each atom to the center of the cage equals to 3.956 \AA, which is in
excellent agreement with the value obtained by Zope \emph{et al.}  \cite{zbpd05}, 3.94 \AA. 
The HOMO-LUMO gap obtained by us, 4.7 eV,  also agrees very well with the Zope \emph{et al.} findings, 5.0 eV.

For the cubic boron nitride, we have obtained a value for the lattice parameter of 3.643 \AA, which differs
in less than 1$\% $ to the experimental value of 3.6157 \AA~ \cite{kwjc89}, and to other DFT-GGA calculations, 3.625 \AA~ \cite{zh01}.
Our methodology predicts a wide band gap for the cubic boron nitride, 4.45 eV, which accurately 
matches other DFT-GGA calculations, 4.43 eV  \cite{zh01}, and is consistent with the experimental result, 6.4 eV  \cite{c74}.

For a single FeO molecule, we have obtained a bond length of 1.620 \AA, in accordance with other theoretical result, 1.606 \AA ~
\cite{wssy99}, and with the experimental value, 1.618 \AA ~\cite{cgm91}. 
We have also performed calculations on FeO bulk obtaining the Fe-O distance of 2.15 \AA, 
which fairly reproduces the experimental results, 2.154  and 2.165 \AA~ \cite{fhvs02}.

\section{Results}

We have firstly studied the interaction between a single FeO molecule and a (10,0) single wall BN nanotube.
We have found that the FeO molecule strongly interacts with the tube, with binding energy of 1.7 eV. In the 
optimized structure the Fe atom binds to the N atom with bond length of 
$\sim$2.1~\AA~ and the O atom binds to the B atom with bond length of 1.55 \AA. 
The presence of the FeO molecule leads to distortions in the bonds of the nanotube. 
Such a bond geometry in the tube is similar to that of a FeO molecule interacting with the octahedral $B_{36}N_{36}$ 
fullerene obtained in a previous work \cite{bmc07}. However, the FeO molecule does not break any B-N
bond of the tube as seen in the fullerene case. The B-N bond length increases from 1.46 \AA~ to 1.62 \AA~ in the 
BN nanotube whereas it increases from 1.46 \AA~
 to 3.26 \AA~ in  the $B_{36}N_{36}$ fullerene due to the presence of an adsorbed FeO molecule in the system. 
 Such a difference can be account to the curvature of the edge of the octahedral fullerene, which is greater than the curvature of the tube, 
 making the edge of the fullerene more reactive. Indeed, the value of the binding energy between the FeO molecule and the 
 BN nanotube (1.7~eV) is much smaller than the one between the FeO molecule and the BN fullerene (3.92 eV). Such an analysis 
 is consistent with other theoretical works in which the curvature due radial compression in the tube increases the tube 
 reactivity \cite{zg09,amc08}. Similarly to the case of the fullerene, the magnetic moment of the Fe molecule does not change 
 due to its interaction with the tube and the total magnetic moment of the optimized structure is 4.0 
 $\mu_\mathrm{Bohr}$.

\begin{figure}[htbp]
   \centering
   \includegraphics[scale=0.4,clip=true]{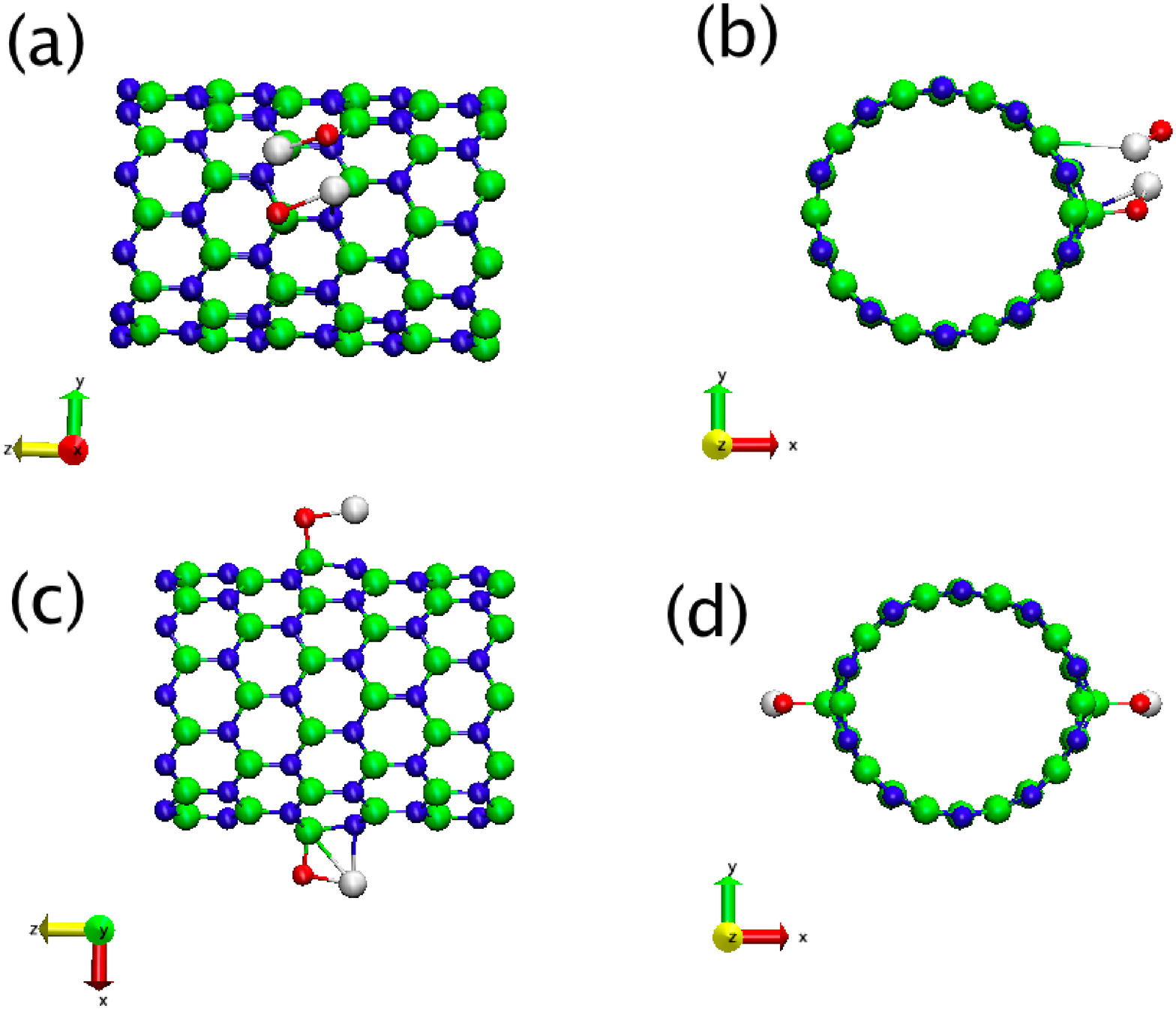}
   \caption{Different point of views for the BN nanotube with two bound FeO molecules. (a) Two bound neighboring 
   FeO molecules viewed along the $x$- and (b) $z$-direction and (c) two distant bound FeO molecules viewed along 
   the $y$- and (d) $z$-axis. The configuration showed in panels (c) and (d) is  3.46 eV more energetic than the one showed in (a) and (b).
   Thus it is energetically preferable to the system to have additional FeO molecules binding nearby already bounded FeO molecules.}
   \label{fig:tubopares}
\end{figure}

In order to address the question of how  additional molecules would bind to the BN tube %we included one more FeO molecule to the system. 
we have considered two binding sites for an additional molecule: (i) neighboring  
the first molecule, as shown in Fig. \ref{fig:tubopares} (a) and (b), and (ii) far way from it, as can be seen from 
Fig. \ref{fig:tubopares}(c) and (d). We have found that is energetically favorable for the second molecule 
to bind sites nearby to the first adsorbed molecule. 
The total energy of the system shown in Fig. \ref{fig:tubopares} (a) and (b) is 3.46 eV lower than the total energy of the system shown in 
Fig. \ref{fig:tubopares} (c) and (d). In fact, when FeO molecules bind to each other the coordination number of the 
FeO structure approaches to the six neighbors atoms found in the FeO bulk,  increasing then the energetic stability of the system.
The magnetic moment of the FeO molecules does not change due mutual interaction and the total magnetic 
moment of the structure shown in Fig. \ref{fig:tubopares} (a) and (b)  is 8.0 $\mu_\mathrm{Bohr}$.
%In the structure shown in Fig.\ref{fig:tubopares} (a) and (b), each Fe atom binds a N (the bond lelengths ?? \AA) and the O  

\begin{figure}[htbp]
   \centering
   \includegraphics[scale=0.4,clip=true]{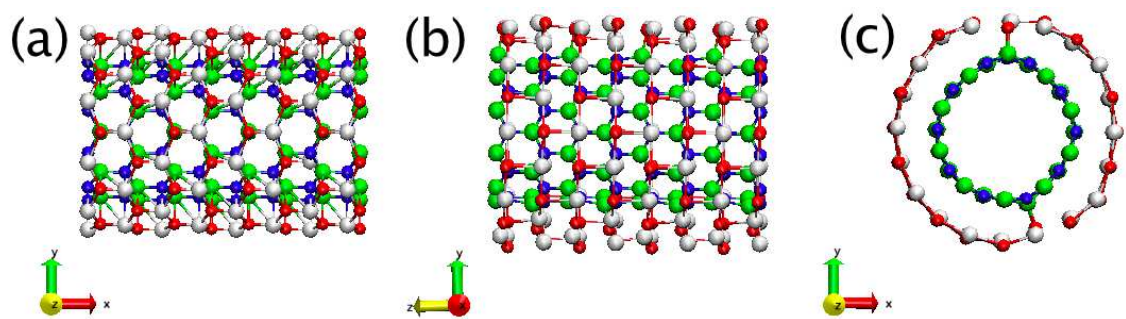}
   \caption{(a) Quasi-hexagonal initial configuration for the FeO monolayer covering the BN nanotube. The system was
   allowed to relax and the minimum energy configuration is as shown in panel (b). In panel (c) is shown the
   same structure as in (b), but along the $z$-direction instead.}
   \label{fig:tubocoberto}
\end{figure}

The results previously described suggest that additional FeO molecules tend to bind to the BN tube nearby to already 
bounded molecules. If the number of FeO molecules is sufficiently large, the tube could be fully covered by a FeO 
layer forming a magnetic nanotube with only 1.28~nm in diameter (see Fig. \ref{fig:tubocoberto}). 
This case is particularly interesting since magnetic nanotubes of iron oxide, Fe$_{2}$O$_{3}$, have 
presented spin filter effect and anomalous magnetoresistence \cite{llxz06}. 
Fig. \ref{fig:tubocoberto} shows the geometry of the (10,0) BN nanotube covered by a layer of FeO proposed in this study. 
We have considered a initial geometry for the FeO layer  as quasi-hexagonal structure in which each Fe atom binds to a N atom 
and to three O atoms [see Fig. \ref{fig:tubocoberto} (a)]. 
Nevertheless, the optimization of the structure increases the coordination number of each Fe atom. 
After the geometry optimization, Fe and O atoms form a square lattice in which each Fe atom binds to four O atoms, 
as one see from Fig. \ref{fig:tubocoberto} (b). 
Such a layer presents similarities to the FeO bulk. Besides the angle formed by adjacent Fe-O bonds, which 
is about 90 degrees, the average bond length between Fe and O atoms, 2.11~\AA, is also similar to that one calculated for the 
FeO bulk, 2.15 \AA. Such results indicate that several layers of FeO on a BN nanotube would
present a local crystal structure similar to that one of the FeO bulk. However, 
due the curvature of the FeO layers, such 
nanotubes may present novel properties compared with those of the FeO bulk.

\begin{figure}[htbp]
   \centering
   \includegraphics[scale=0.45,clip=true]{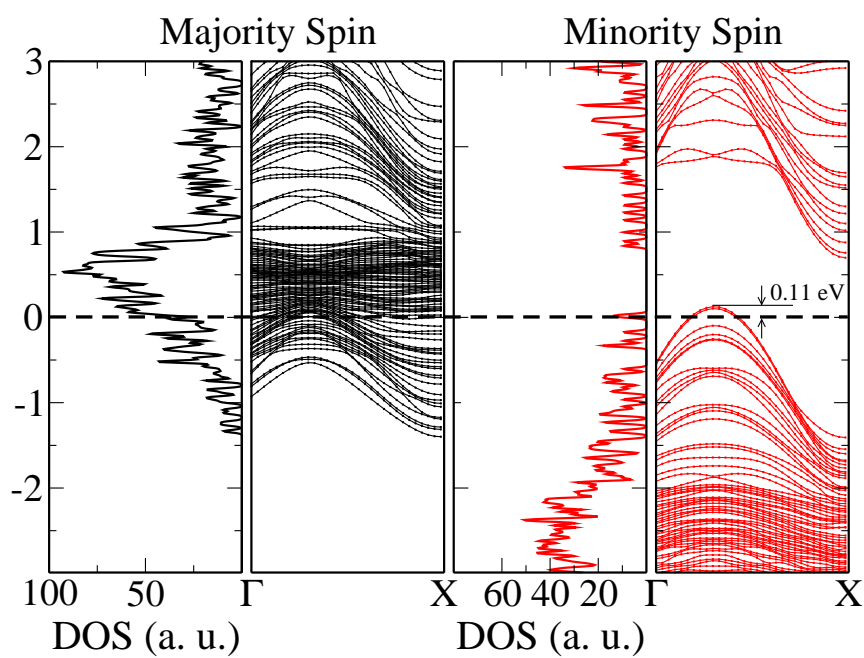}
   \caption{Density of States and band structure of FeO Covered (10,0) BN nanotube calculated using a 80 atoms unit cell with 
   (1$\times$1$\times$6) Monkhorst-Pack grid. The Fermi level is set to zero in this figure. The band structure for minority spin shows that the FeO covered nanotube is a semi-half-metal with a band gap of 0.5~eV at roughly 0.11~eV above the Fermi level.}
   \label{fig:bandas}
\end{figure}

We have observed that the system formed by a single FeO layer on the BN tube 
presents electronic properties which contrast with those of the FeO bulk. 
Fig. \ref{fig:bandas} shows the band structure and the density of states of the system shown in 
Fig.\ref{fig:tubocoberto} for the two spin states. Consistently with experimental results \cite{llxz06}, 
the band structure for the minority spin shows a band gap just above the Fermi level and therefore the system may 
become a half-metal if a small change in the Fermi level (about 0.11 eV) is induced. 
As it was mentioned by Xiang \emph{et al} \cite{xyhz05}, such a change in the Fermi level 
could be induced by applying a gate voltage in a MOSFET-like system. 
The observed semi-half-metallicity in BN tubes covered by FeO  make it highly suitable to be the 
building blocks for spin-filter devices, just as the case of iron oxides nanotubes seen in the work of Liao \emph{et al.} \cite{llxz06}. 
We have found that the
total magnetic moment per Fe atom, 3.94 $\mu_\mathrm{Bohr}$, 
decreases in comparison to that of small FeO concentration 4.00 $\mu_\mathrm{Bohr}$.

\section{Conclusions}

In summary, our first-principles calculations show that FeO and BN nanotubes
may form stable structures. We found a high value of the binding energy for the FeO molecule
adsorbed in a (10,0) BN nanotube, 1.7~eV. We also noticed that it is energetically favorable
for an additional FeO molecule to aggregate nearby adsorbed molecules instead of bind
sites away from it. In case of small FeO concentrations, the total magnetic moment per unit cell
is the magnetic moment of the Fe atom (4$\mu_\mathrm{Bohr}$) times the number of Fe atoms in 
the unit cell. In case of a layer of FeO molecules covering the BN nanotube, the total magnetic
moment in the unit cell decreases to 3.94$\mu_\mathrm{Bohr}$. 
The BN nanotube covered with a single layer of FeO molecules is a semi-half-metal which 
may become a half-metal if a small change in the Fermi level is induced.

\section{Acknowledgements}

This work was partially supported by the Brazilian
agencies CNPq and FAPEMIG.

\bibliography{SpinFilter25may}

\pagebreak

%\begin{figure}[htbp]
  % \centering
   %\includegraphics[scale=0.4, clip=true]{Energy.eps}
  % \caption{Total energy of the the system against the distance of the Fe atom of a single FeO molecule to the tube wall. We see that
 %  the bond length between the Fe and N atoms is $\sim$2.1 \AA. The FeO molecule was initially placed at a distance of 5.0 \AA~ to the tube wall  and the structure of the system was allowed to relax. The dots correspond to each step of the minimization.}
 %  \label{fig:energy}
%\end{figure}

%\pagebreak

\pagebreak

\pagebreak

  \end{document}